\documentclass[fdp,fleqn]{w-art}
\usepackage{times}
\usepackage{w-thm}
\usepackage{amssymb}
\usepackage{bbm}
\usepackage[]{graphicx}

\newcommand{\SU}[1]{\ensuremath{\mathrm{SU}(#1)}}
\newcommand{\U}[1]{\ensuremath{\mathrm{U}(#1)}}
\newcommand{\Z}[1]{\ensuremath{\mathbbm{Z}_{#1}}} 

\newcommand{\I}{\mathrm{i}}
\begin{document}
\pagespan{1}{}

\title[The $\mu$--problem, the NMSSM and string theory]{The $\mu$--problem, the NMSSM and string theory}

\author[S. Ramos-S\'anchez]{Sa\'ul Ramos-S\'anchez%
  \footnote{Author E-mail:~\textsf{ramos@mail.desy.de}}
\address{Deutsches Elektronen-Synchrotron DESY, Hamburg, Germany}}

\begin{abstract}
  We discuss recent developments on the solution of the so-called supersymmetric $\mu$--problem in the
  context of heterotic orbifolds. In particular, an approximate $R$ symmetry can induce an admissible
  $\mu$--term in Minkowski vacua of orbifold models with the MSSM spectrum. A natural solution to the
  $\mu$--problem is also offered by explicit string-derived NMSSMs. These setups help avoid the
  fine-tuning of the MSSM.
\end{abstract}
\maketitle                   

\section{Introduction}

In supersymmetric extensions of the standard model (SSM), the mixed mass--term of the up and down Higgs
doublets, $\mu H_u H_d\subset W$, must be nonvanishing in order to avoid an undesirable massless
axion~\cite{Kim:1983dt}.  Furthermore, standard model (SM) phenomenology constrains $\mu$ to be of the order
of the soft masses. On the other hand, the most appealing feature of all SSMs is that they provide a valid
description up to a scale as large as $M_{GUT}$ or $M_{Pl}$. This triggers a naturalness issue
traditionally called the $\mu$--problem: why does the supersymmetric mass scale $\mu$ know about the
scale of supersymmetry (SUSY) breakdown?  Stated differently: where does the hierarchy $\mu\ll M_{GUT},\,M_{Pl}$
come from?

Although in the minimal SSM (the MSSM) the $\mu$--term was originally put by hand, it has been shown that
breaking supergravity (SUGRA) can induce a $\mu$--term of the correct order if one includes radiative
corrections~\cite{Hall:1983iz,Ohta:1982wn} or if one assumes a special Higgs--dependent structure of the K\"ahler
potential~\cite{Giudice:1988yz}. Explaining why $\mu$ vanishes before SUGRA breakdown might require
additional symmetries in the theory, such as Peccei--Quinn (PQ) or $R$ symmetries, which would eventually
also fix the so-called strong CP problem.

A perhaps more elegant solution to the $\mu$--problem is provided by the next-to-minimal SSM (the NMSSM)
\cite{Nilles:1982dy,Frere:1983ag} (see~\cite{Ellwanger:2009dp} for a recent review). In the NMSSM,
the $\mu$--term is generated via the introduction of a singlet superfield $S$ with the coupling $\lambda
S H_u H_d \subset W$, where $\lambda$ is a dimensionless parameter. Provided that $S$ remains
massless in the supersymmetric theory, $S$ naturally acquires a vacuum expectation value (vev) of the order of
the SUSY breaking scale, thereby giving rise to an effective $\mu$--term of the correct size. 

A related matter is the question of how much fine-tuning is required in order for a SSM to accommodate
the LEP Higgs bound. In the MSSM, satisfying this bound necessitates large radiative corrections to the
tree--level prediction for the SM--like Higgs boson $h$ ($m_h\leq m_Z$). This can be achieved only if the
superpartner masses are very large (about 1.3 TeV) which would imply considerable fine-tuning of the MSSM
soft terms. In the NMSSM the situation is better due to the existence of a light pseudoscalar $a$. $h$
decays predominantly into pairs of $a$'s and these subsequently decay into taus or light
quarks~\cite{Dermisek:2005ar}. For such final states, the lower LEP bound on the Higgs mass relaxes and can in
some cases be as low as 90 GeV. Consequently, the superpartners are not required to be very heavy for the
Higgs mass to comply with this bound, and the fine-tuning problem of the MSSM can be avoided.

To single out one solution to these riddles, a top--down approach might be of great help. String theory
is perhaps the best known candidate to provide some clues about physics from very large energies down to
the electroweak scale. Therefore, promising stringy constructions can
reveal the correct solution or provide new suitable tools for the resolution of e.g. the problems
described before. This approach has the advantage that, unlike the field-theoretic solutions just described, string-derived
models are believed to be ultraviolet complete and all their admissible interactions and matter content
are fixed by the theory itself. This implies that any new feature of phenomenologically acceptable string-derived models
can be considered a prediction. The challenge is then to build such acceptable models.

The main hurdle to addressing low energy physics from string theory is that it predicts the existence of
six additional spatial dimensions on top of the four--dimensional space--time of our everyday life
experience. To explain this discrepancy, it is typically argued that the extra coordinates are compact
and too small to be detected at currently achievable energies, or that they are ``invisible'' to us
because our experiences are limited to a four--dimensional subspace or brane in which we happen to live.
Both interpretations have been explored and led to semi--realistic
models~\cite{Ibanez:1987sn,Font:1988mm,Casas:1988se,Dijkstra:2004cc,Blumenhagen:2006ux,Faraggi:2006bc,Gmeiner:2008xq}.
Among them, there is a relatively small number of constructions with the exact MSSM spectrum
\cite{Buchmuller:2005jr,Lebedev:2006kn,Kim:2006hv,Blaszczyk:2009in,Bouchard:2005ag,Braun:2005nv,Anderson:2009mh}.

Since the first studies, orbifold compactifications of the heterotic string have demonstrated to be
optimal candidates to accommodate the properties of our universe. They are four-dimensional compact
spaces divided by a discrete symmetry, what gives rise to a finite number of curvature singularities to
which the matter states are attached. Particularly interesting are the orbifold models of the \Z6--II
heterotic mini-landscape~\cite{Lebedev:2006kn,Lebedev:2008un}. Remarkably, it has been found that
$\mathcal{O}(100)$ models in this scenario comply with the requirements of displaying the matter spectrum
of the MSSM and gauge unification. What is more surprising is that models satisfying these requirements
are automatically endowed with further appealing features, such as matter parity, low-scale SUSY
breakdown, gauge-top unification, seesaw neutrino masses, flavor symmetries and potentially realistic
fermion masses (see e.g.~\cite{Nilles:2008gq} for a review). In this paper we show how also the
$\mu$--problem is naturally solved in these constructions.

There are several approaches to address the $\mu$--problem in string constructions. For instance,
admissible effective $\mu$--terms can be generated by (i) stringy instantons~\cite{Green:2009mx}, (ii)
string threshold corrections, (iii) particular structures of the K\"ahler potential, (iv) nonstandard
supergravity interactions~\cite{Antoniadis:1994hg}, and (v) explicit superpotential
masses~\cite{Casas:1992mk,Buchmuller:2008uq,Kappl:2008ie}.  In the remainder of this paper, we concentrate
on the latter approach and address the origin of an admissible $\mu$--term in 
orbifold (N)MSSMs.

Our discussion is organized as follows. In sec.~\ref{sec:muInmssm}, we study how a suppressed $\mu$--term
appears in Minkowski vacua as a consequence of an approximate $\U1_R$ symmetry.  In sec.~\ref{sec:nmssm},
we briefly address the main properties of the NMSSM candidates arising from heterotic orbifolds and
discuss the features of an orbifold example. Finally, sec.~\ref{sec:conclusions} is devoted to
some final remarks.

\section{Solving the $\mu$--problem in stringy MSSMs~\cite{Kappl:2008ie}}
\label{sec:muInmssm}

In ref.~\cite{Casas:1992mk}, it is argued that an elegant solution to the $\mu$--problem can be achieved
under the assumption that the operator $H_u H_d$ is vectorlike w.r.t. all gauge and string symmetries. In fact, in
many of the  mini-landscape models and, in particular in the benchmark model 1 of
ref.~\cite{Lebedev:2007hv}, this is true. As a consequence, any superpotential term $W_0^j$ allowed by
string selection rules will also couple to $H_u H_d$. The superpotential can then be written as
$W=W_0+\alpha W_0 H_u H_d$, with $W_0=\sum_j W_0^j$ being a polynomial on the singlet fields $s_i$. An
effective $\mu=\alpha \langle W_0\rangle$ appears once the $s_i$'s develop vevs. However, from a top-down
perspective, $\langle W_0\rangle\ll M_{Pl}$ seems rather ad hoc. In the following, we discuss a
natural explanation of this hierarchy.

Clearly, $\langle W_0\rangle$ depends on the vevs of the singlets, which are subject to the
SUSY constraints $F=D=0$. 
The requirement to cancel the  Fayet-Iliopoulos D--term, commonly present in heterotic
orbifolds~\cite{Dine:1987xk}, does not fix the size of the vevs $\langle s_i\rangle$, but it introduces in
the problem a new scale $\sqrt{\xi}\sim0.1$ in Planck units. This results on singlet vevs of the same
order,\footnote{In some cases, $\mathcal{O}(1)$ vevs are also possible.} which are in general fixed by $F=0$.

Remarkably, apart from the features discussed above, it was noted that vanishing F--terms imply that $W$
(truncated at order $N$) cancels term by term. The reason was found to be that, when this happens, $W$ is
endowed with an approximate global $\U1_R$ symmetry. This can be seen as follows. Under the $R$ symmetry,
$W$ and $s_i$ transform respectively as
\begin{equation}
  \label{eq:Rtrafo}
W\rightarrow e^{2\I\beta}W \qquad{\rm and }\qquad s_i\rightarrow s_i'=e^{\I r_i\beta}s_i\,.  
\end{equation}
On the other hand, an infinitesimal $\U1_R$ transformation of $W$ yields 
\begin{equation}
  \label{eq:infRtrafo}
  W(s_i)\rightarrow W(s_i')=W(s_i)+\sum_j F_j \Delta s_j  \,,
\end{equation}
which reduces to $W(s_i)$ in a SUSY-preserving vacua (i.e. $\langle F_i\rangle=0$). This is consistent
only if $\langle W\rangle=\langle W_0\rangle$ vanishes. As a conclusion, we find that a supersymmetric
theory with a $\U1_R$ symmetry yields naturally Minkowski vacua.\footnote{Note that this also holds in
  SUGRA, for $D_i W_0=0$.}

The $\U1_R$ we observe in our models is a low--energy realization of exact discrete symmetries of stringy
origin. Then $\U1_R$ has to be broken explicitly by terms of higher order $>N$.This has two
advantages. First, if the $R$ symmetry is broken at order $N+1$, $\langle W\rangle=0$ is no longer
protected and its nonvanishing value is proportional to $\langle s_i\rangle^{N+1}$. In
supergravity theories, it follows then that the gravitino mass is $m_{3/2}\sim \langle
s_i\rangle^{N+1}$. Second, the pseudo-Goldstone boson generated by the breaking of the $R$ symmetry
acquires a mass of order $\langle s_i\rangle^{N-1}$, i.e. enhanced w.r.t. the gravitino mass and thus
consistent with current bounds.

As an example, let us consider the benchmark model 1 of ref.~\cite{Lebedev:2007hv}. It turns out that
this model is furnished with an approximate $\U1_R$ symmetry which is preserved up to order 10. (However,
in other models similar symmetries are unbroken up to orders as high as 26.) Since $\langle
s_i\rangle\sim0.1$ then Minkowski vacua with $\mu\sim\langle W\rangle \sim \mathcal{O}(10^{-11})$ or
smaller emerge naturally from promising heterotic orbifolds.

Notice that this solution to the $\mu$--problem is only an application of a more
interesting finding: the scheme described above can explain the origin of large hierarchies in a natural
way. This is a major achievement considering that precisely this question is one of most intriguing puzzles of
contemporary physics. The hierarchies generated in this way are important to solve many phenomenological
issues, such as moduli stabilization~\cite{Dundee:2010sb} and the strong CP problem~\cite{Choi:2009jt}.
For a detailed discussion, see ref.~\cite{Brummer:2010fr}.

\section{The NMSSM from string theory~\cite{Lebedev:2009ag}}
\label{sec:nmssm}

It is known that in heterotic orbifolds, the $\mu$--term does not arise at trilinear
level~\cite{Antoniadis:1994hg}. However, it can appear effectively from couplings of the Higgs pair 
to some SM singlets, which are quite abundant in the mini-landscape models. This motivates the study 
of the NMSSM in string models.

Apart from the standard couplings and matter content of the MSSM, the (\Z3-invariant) NMSSM includes a
massless (at the string level) singlet $S$ with the following superpotential contributions
\begin{equation}
\label{nmssm}
W= \lambda S H_u H_d + \tfrac13\kappa S^3 \;,
\end{equation}
We are assuming that SUSY is not broken by the F--term of $S$ (in the limit $\langle S \rangle
\rightarrow 0$) and thus the ``tadpole'' term linear in $S$ is also negligible. In what follows, we will
assume that all relevant soft parameters, $m^2_{H_u},m^2_{H_u},m^2_S,A_\lambda,A_\kappa$, are of
the electroweak (EW) size whereas $\lambda$ and $\kappa$ can take arbitrary values.  After SUSY breaking,
including soft terms, the potential for (the real part of) the scalar component of $S$, denoted by $s$,
is given by
\begin{equation}
  \label{eq:gralVofs}
  V(s)=-2\lambda A_\lambda v_u v_d s + m_S^2 s^2 +(\kappa s^2- \lambda v_u v_d)^2 +(\lambda v_d s)^2
  +\tfrac23\kappa A_\kappa s^3\,,
\end{equation}
where $v_{u,d}=\langle H_{u,d}\rangle$.

In heterotic orbifolds, $\lambda $ and $\kappa$ are effective couplings of the form  
\begin{equation}
\lambda = {\rm const} + \langle s_{a_1} s_{a_2}\cdots \rangle ~,\qquad 
\kappa  =  \langle s_{b_1} s_{b_2}\cdots \rangle \;, \label{c}  
\end{equation}
where, as in sec.~\ref{sec:muInmssm}, $s_i$ are SM singlets attaining $\mathcal{O}(0.1)$ vevs in Planck
units.  The SM singlet $S$ typically comes from the gauge sector and thus carries charges under e.g.
some additional unbroken gauge \U1s. These symmetries are violated by $S^3$ and only after they get
broken spontaneously, is this effective interaction allowed.  As a result, $\kappa$ is suppressed by the
SM singlet vevs $\langle s_{b_1} s_{b_2}\cdots \rangle$. Note that if $S$ is a modulus, it is neutral
under gauge symmetries but its interactions are Planck suppressed and we arrive to the same conclusion.
In contrast, a coupling among three different fields can be allowed already at the trilinear level, hence
the ``const'' term in eq.~\eqref{c}.\footnote{Note that $\lambda$ is suppressed if e.g. (as required in
the previous section) $H_u H_d$ is vectorlike, unless $S$ is a gauge singlet. We have verified that such
singlets do not appear in the mini-landscape models.}

Therefore, typically $\kappa \ll 1$ while $\lambda$ can be order one. If the ``const'' term in
eq.~\eqref{c} vanishes due to string selection rules, then $\lambda$ is also suppressed. We thus are led
to two distinct versions of the NMSSM: the ``decoupling'' ($\lambda,\kappa \ll 1$) and the Peccei-Quinn
scenarios ($\kappa \ll 1$), which we now discuss.

1) {\it Decoupling limit}.  For $\lambda,\kappa \ll 1$, the singlet essentially decouples and the NMSSM
degenerates into a version of the MSSM, albeit with modifications in the neutralino sector.  The dominant
terms for large $s$ in the potential~\eqref{eq:gralVofs} are
\begin{equation}
  \label{eq:decVofs}
  V(s) \sim m_S^2 s^2 + \tfrac23 \kappa A_\kappa s^3 + \kappa^2 s^4\,.
\end{equation}
For $A_\kappa^2 \geq 8 m_S^2$, there is a local minimum at $s \simeq \frac{1}{\kappa}(- A_\kappa +
\sqrt{A_\kappa^2 -8 m_S^2})$. In the decoupling limit, $s$ can take very large vevs and still satisfy
the chargino mass bound, $\lambda s \sim~ $EW, thus solving the $\mu$--problem. We then have $s
\sim \frac{\rm EW}{\kappa} \sim \frac{\rm EW}{\lambda}$. The
difference from the MSSM resides in the neutralino sector: the fermionic component of $S$ has mass
$2\kappa s$ and can be the LSP.  The NLSP decays are then suppressed by the small coupling $\lambda$
leading to its long lifetime with characteristic signatures such as displaced vertices
\cite{Ellwanger:1997jj}.

2) {\it Peccei--Quinn limit}. For $\kappa \ll 1$ \cite{Miller:2003ay}, the model possesses an approximate
PQ symmetry $H_{u,d} \rightarrow e^{i \alpha} H_{u,d} $, $S \rightarrow e^{-2 i \alpha} S$. Spontaneous
breaking of this symmetry generates a pseudo--Goldstone boson (axion). The composition of this state is
given by
\begin{equation}
 a_{\rm PQ} =~(v \sin 2\beta~A   -2 s~ S_I)\Big/\sqrt{v^2 \sin^2 2\beta +4 s^2}\;,
\end{equation}
where $A= \cos\beta ~H_{uI}+\sin\beta~H_{d I}$,\ and $S_{I}, H_{uI},H_{dI}$ are defined by $f_I \equiv
\sqrt{2}~ {\rm Im}(f-\langle f \rangle) $. As usual, $\tan\beta=v_u/v_d$ and $v=\sqrt{v_u^2 +v_d^2}=174$
GeV. Since the PQ is only slightly broken, $a_{\rm PQ}$ can get a rather small mass-square of order $3\kappa s
A_k$. 

\begin{figure}[!t!]
\begin{minipage}{0.47\textwidth}
(a)\\
$\vcenter{\centerline{\hbox{\includegraphics[width=.6\textwidth]{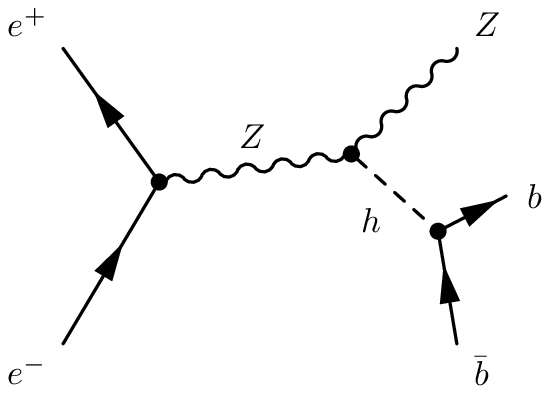}}}}$
\end{minipage}
\hskip 2mm
\begin{minipage}{0.47\textwidth}
(b)\\
$\vcenter{\centerline{\hbox{\includegraphics[width=.75\textwidth]{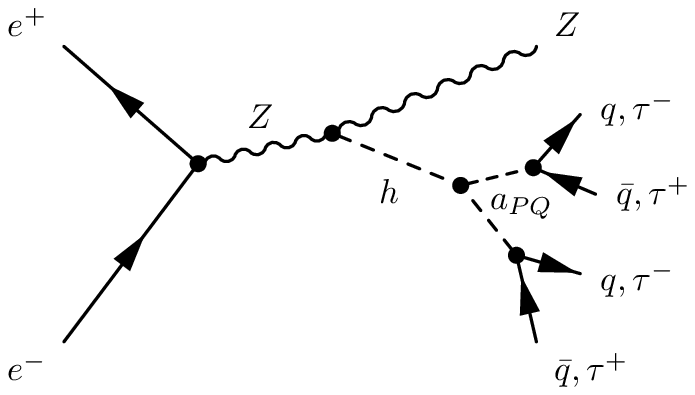}}}}$
\end{minipage}
\caption{Possible Higgs detection channels. In the case (a) a signal with $m_h\lesssim114$ GeV 
is excluded by LEP data. The LEP bound relaxes in the case (b) admitting $m_h$ around 105 GeV (90 GeV)
for $\tau$'s (light quarks) in the final state.
\label{fig:SSMhiggsdecay}}
\end{figure}

As in the previous case, the $\mu$--problem is solved because $s$ can be stabilized at values around (or
larger than) the EW scale.  In addition, the presence of a light axion--like state can be relevant to the
MSSM fine-tuning problem \cite{Dermisek:2005ar}.  Typically, $s\gg v \sin 2\beta$, so that the axion is
predominantly an EW singlet.  Its couplings to quarks and gauge bosons are suppressed, but the coupling
to the Higgs bosons is significant. Thus the SM-like Higgs $h$ can decay into pairs of $ a_{\rm PQ}$
which would subsequently decay into 4 fermions. If $m_{a_{\rm PQ}}< 2 m_b$, the dominant decay channel at
LEP $h\rightarrow b\bar{b}$ would be overwhelmed by $h\rightarrow 2 a_{\rm PQ} \rightarrow 4 \tau ~(4 q)$,
with $q$ denoting light quarks (see fig.~\ref{fig:SSMhiggsdecay}).  Under these conditions, the LEP bound
on the Higgs mass relaxes to about 105 GeV for the final state taus and 90 GeV for the final state
quarks~\cite{Dermisek:2005ar}.  This ameliorates (or even eliminates) the MSSM fine-tuning problem since
the superpartners are not required to be very heavy to accommodate the LEP Higgs bound.

\vskip 2mm

To obtain an example of the NMSSM from the mini--landscape models, one needs to impose the extra
requirement that there exist at least one massless singlet which couples to the Higgs pair.  This
condition turns out to be very restrictive. In particular, among the models with 2 Wilson lines
of ref.~\cite{Lebedev:2006kn} there are no NMSSM examples because all SM singlets are decoupled once the
exotics acquire large masses. However, many models with 3 Wilson lines of ref.~\cite{Lebedev:2008un}
lead to the most appealing scenario: the PQ limit of the NMSSM.

In the particular model discussed in~\cite{Lebedev:2009ag}, the unbroken gauge group after solving the
SUSY constraints is $\SU3_C\times\SU2_L\times\U1_Y\times[\SU6\times\U1]$, where the gauge
factors in parenthesis are hidden, in the sense that this sector communicates to the SM sector
only through gravitational interactions. The matter content
includes three generations of quarks and leptons, very heavy vectorlike exotics, and an additional
massless singlet $S$ with the couplings
\begin{equation}
  \label{eq:nmssmexample}
  W= S H_u H_d + \tfrac13 \langle s_i\rangle^6 S^3 \;.
\end{equation}
For $\langle s_i\rangle<1$, $\kappa \sim \langle s_i\rangle^6\ll 1$ and the system has an approximate PQ
symmetry, whose spontaneous breaking results in a light pseudoscalar state $a_{\rm PQ}$.  Its mass
depends on the order of the allowed coupling as well as the exact value of $\langle s_i\rangle$ and can
be light enough to be relevant to the MSSM fine-tuning problem. In this model choosing 
$A_\kappa,A_\lambda\sim10^2\ {\rm GeV},1<\tan\beta<10$ and minimizing the potential~\eqref{eq:gralVofs} leads to 
$m_{a_{\rm PQ}}\sim 100$ MeV and a $\mu$--term of about the right size. 

\section{Final remarks}
\label{sec:conclusions}

The solutions to the $\mu$--problem addressed here in the context of string-derived models rely on
supplementary symmetries of the theory. These symmetries are frequently artificial elements from the low energy
perspective. However, particularly in heterotic orbifolds, they are consequences of the stringy UV completion of these effective theories.
Unlike in field theories, no symmetry is put by hand. The $\U1_R$ symmetry necessary to solve the
$\mu$--problem in sec.~\ref{sec:muInmssm} and the PQ symmetry that yields $\kappa\ll 1$ in the
string-derived NMSSM of sec.~\ref{sec:nmssm} are the result of the string selection rules together with
our vacuum selection.  Also, in the string NMSSM the absence or suppression of the $S^3$ term has to do
with the fact that $S$ is charged under additional gauge symmetries. Similarly, the vectorlikeness of the
operator $H_u H_d$ is a result of the stringy origin of the symmetries of the model.

The explicit breaking of the approximate symmetries we have studied cures some of their usual problems when they are exact.
For instance, the $\Z3$ symmetry of the NMSSM is broken by a small supersymmetric mass
term for the singlet. This helps avoid cosmological problems, such as domain walls. Further, the explicit
breaking of the $\U1_R$ needed for solving the $\mu$--problem in the MSSM renders heavy an otherwise
massless Goldstone boson of the theory.

Let us conclude by noting that the models we discussed here are also embedded with seesaw masses,
low-energy SUSY breaking, nontrivial quark and lepton masses, order one top Yukawa coupling and other attractive features. It seems
then that the conjecture that our universe might well be described by compact singular spaces rather than
smooth manifolds~\cite{Nilles:2009yd} should not be ignored and cannot be ruled out.

\begin{acknowledgement}
  It is a pleasure to thank to O.~Lebedev, H.P.~Nilles, M.~Ratz and P.~Vaudrevange for
  enjoyable collaborations and many insightful discussions.
\end{acknowledgement}

\addcontentsline{toc}{chapter}{Bibliography}

\providecommand{\bysame}{\leavevmode\hbox to3em{\hrulefill}\thinspace}

\end{document}